\documentclass[12pt]{article}
\pdfoutput=1
\usepackage{jheppub}
\usepackage{feynmp}
\newcommand{\be}{\begin{equation}}
\newcommand{\ee}{\end{equation}}
\newcommand{\bea}{\begin{align}}
\newcommand{\eea}{\end{align}}
\begin{document}
\subheader{SU-ITP-12/29}
\title{Gauge Invariant Computable Quantities In Timelike Liouville Theory}
\author[a]{Jonathan Maltz}
\affiliation[a]{Stanford Institute for Theoretical Physics and Department of Physics,\\ Stanford University, Stanford, CA, 94305}
\emailAdd{jdmaltz@Stanford.edu}
 \setlength{\unitlength}{1mm}
\def\Re{\mathrm{Re}}
\def\Im{\mathrm{Im}} 
 \def\M{{\mathcal M}}
\def\hat{\widehat}
 \def\tilde{\widetilde}
\def\V{{\mathcal V}}
\def\C{{\mathbb C}}
\def\F{{\mathcal F}}
\def\O{{\mathcal O}} 
\def\Z{{\mathbb Z}}
\def\Tr{{\mathrm{Tr}}}
\def\R{{\mathbb R}}
\def\RR{{\mathcal R}}
\def\S{{\mathbb S}}
\def\I{{\mathcal I}}
\def\D{{\mathcal D}}
\def\A{{\mathcal A}}
\def\bar{\overline}
\def\T{{\mathcal T}}
\def\J{{\mathcal J}}
\def\b{{\hat{b}}}
\def\tilde{\widetilde}
\def\P{{\text{LegendreP}}}
\def\Q{{\text{LegendreQ}}}
\def\p{{\delta\varphi}}
\def\t{{\delta\theta}}
\abstract{Timelike Liouville theory admits the sphere $\mathbb{S}^{2}$ as a real saddle point, about which quantum fluctuations can occur. An issue occurs when computing the expectation values of specific types of quantities, like the distance between points. The problem being that the gauge redundancy of the path integral over metrics is not completely fixed even after fixing to conformal gauge by imposing $g_{\mu\nu} = e^{2\b\phi}\tilde{g}_{\mu\nu}$, where $\phi$ is the Liouville field and $\tilde{g}_{\mu\nu}$  is a reference metric. The physical metric $g_{\mu\nu}$, and therefore the path integral over metrics still possesses a gauge redundancy due to invariance under $SL_{2}(\mathbb{C})$ coordinate transformations of the reference coordinates. This zero mode of the action must be dealt with before a perturbative analysis can be made.  

This paper shows that after fixing to conformal gauge, the remaining zero mode of the linearized Liouville action due to $SL_{2}(\mathbb{C})$ coordinate transformations can be dealt with by using standard Fadeev-Popov methods. Employing the gauge condition that the ``dipole" of the reference coordinate system is a fixed vector, and then integrating over all values of this dipole vector. The ``dipole" vector referring to how coordinate area is concentrated about the sphere; assuming the sphere is embedded in $\mathbb{R}^{3}$ and centered at the origin, and the coordinate area is thought of as a charge density on the sphere. The vector points along the ray from the origin of $\mathbb{R}^{3}$ to the direction of greatest coordinate area.

 A Green's function is obtained and used to compute the expectation value of the geodesic length between two points on the $\mathbb{S}^{2}$ to second order in the Timelike Liouville coupling $\b$. This quantity doesn't suffer from any power law or logarithmic divergences as a na\"{i}ve power counting argument might suggest.
 }
\maketitle
\section{Introduction}
\hspace{0.25in}Liouville theory has been a useful component in String Theory and 2-D quantum gravity ever since Polyakov introduced it in the context of Non-critical String theory\cite{Polyakov:1981rd}.  A complete list of the applications of Liouville theory is beyond the scope of this paper. Some applications of note include include its use as a non-compact conformal field theory, a model for Higher-dimensional Euclidean gravity, and as a linear dilaton background in String Theory. It is deeply ingrained in proposed holographic duals of de Sitter space and the multi-verse including the conjectured FRW/CFT \cite{Freivogel:2006xu,Susskind:2007pv,Sekino:2009kv,Harlow:2010my}. It has also been found to have a connection to four-dimensional gauge theories with extended supersymmetry \cite{Alday:2009aq}. Work on Liouville theory has yielded exact results from , e.g. the correlation function of three primary operators which is given by the  DOZZ formula \cite{Dorn:1994xn}. Combinatorial approaches have also been made to obtain results in Liouville some examples are \cite{Ambjorn:1995dg,Ambjorn:2011rs,Ambjorn:2012kd}. It has also been used Kaluza-Klein constructions as an explicit model of how spontaneous breaking of space-time translation invariance can lead to compactification of the space-time \cite{PhysRevD.28.2583,PhysRevLett.50.1719}. Recently its path integral properties under analytic continuation have been discussed in\cite{Harlow:2011ny} including the continuation of theory to the Timelike Liouville regime\cite{springerlink:10.1007/s11232-005-0003-3,Harlow:2011ny}. 

      Timelike liouville theory possesses $\mathbb{S}^{2}$ as a real saddle point about which quantum fluctuations can occur. Computing expectation values of fields on this fluctuating geometry involves a path integral over the metric of the geometry. The gauge redundancy of this path integral must be dealt with before meaningful quantities can be computed. The issue that comes up in computing the expectation values of standard classical quantities like the distance between points in this fluctuating geometry is that even after fixing to conformal gauge by imposing  $g_{\mu\nu} = e^{2\b\phi}\tilde{g}_{\mu\nu}$, where $\phi$ is the Liouville field and $\tilde{g}_{\mu\nu}$  is a reference metric of $\mathbb{S}^{2}$, not all the gauge redundancy has been removed. The remaining gauge redundancy is due to $SL_{2}(\mathbb{C})$ which transform the reference coordinates and Liouville field transform nontrivally leaving the physical manifold invariant. This invariance means that until this redundancy is fixed, the integral over metrics is not defined. Computing quantities that depend on the physical points by characterizing them with reference coordinates is not possible because the position of two points on the physical manifold is not uniquely determined by two reference points. A $SL_{2}(\mathbb{C})$ transformation will change the position of the reference points leaving the physical points alone. Computing the distance between the physical points by integrating over the reference points is not defined until the  $SL_{2}(\mathbb{C})$ redundancy is fixed.  In this paper it is shown in a perturbative analysis that after fixing to conformal gauge and expanding about the spherical saddle of the Timelike Liouville field, the remaining zero mode due to the invariance under $SL_{2}(\mathbb{C})$ coordinate transformations of the reference sphere can be dealt with but using standard Fadeev-Popov methods employing the gauge condition that the ``dipole'' of the coordinate system is a fixed vector, and then integrating over all values of this dipole. Dealing with this zero mode means that a Green's function can be obtained  and a pertubative analysis of quantities on spherical geometry under the influence of fluctuations of a semi-classical Timelike Liouville field can be carried out. 

One such quantity is the expectation value of the length of a geodesic on a spherical geometry under the influence of a semi-classical Timelike Liouville field, computed to second order in the Timelike Liouville coupling $\b$. It is shown that this quantity is well defined and doesn't suffer from any power law or logarithmic divergences as a na\"{i}ve power counting argument might suggest.

{\bf{Outline:}} In section \ref{propagator}, a Green's function is obtained by implementing the gauge constraint of fixing the coordinate dipole and integrating over the value of this dipole.
In Section \ref{dist}, the Green's function is employed to compute the expectation value of the separation between two points on sphere under the influence of a fluctuating Timelike Liouville field to second order in the coupling $\b$. Finally in Section \ref{res}, some possible further applications are looked at.

For some modern reviews on Liouville theory the reader is encouraged to look at  \cite{Ginsparg:1993is,Nakayama:2004vk,Teschner:2001rv}, some slightly older reviews include \cite{PhysRevD.26.3517, Zamolodchikov:1995aa,Seiberg:1990eb}. For information on the analytic continuation of Liouville to the Timelike regime the reader is humbly referred to \cite{springerlink:10.1007/s11232-005-0003-3,Harlow:2011ny}.
 
\section{The Gauge Fixed Propagator}\label{propagator}
\hspace{0.25in} When coupling a generic conformal Field to two dimensional gravity, the Liouville action 
\be\label{liouvilleaction}
S_{L} = -\frac{1}{4\pi}\int dx^2 \sqrt{\tilde{g}}\big(\tilde{g}^{ab}\partial_{a}\phi\partial_{b}\phi +Q\tilde{R}\phi + 4\pi\mu e^{2b\phi}\big)
\ee
is obtained after fixing to conformal gauge \cite{David:1988hj,Ginsparg:1993is}\footnote{This means that in the path integral over  metrics a general metric is decomposed into a conformal Liouville factor and a family of conformally inequivalent reference metrics $\tilde{g}_{\mu\nu}$\cite{David:1988hj}. In this paper the only relevant reference geometry is, $\mathbb{S}^{2}$ as higher genus surfaces will not be discussed.}.  (\ref{liouvilleaction}) is invariant under conformal transformations of the coordinates 
\begin{align}\label{anomaly}
z^{\prime} &= w[z]\\
\phi^{\prime}[z^\prime,\bar{z}^{\prime}] &= \phi[z,\bar{z}] - \frac{Q}{2}\log{\Big|\frac{\partial w}{\partial z}\Big|}
\end{align}
with $Q = b + \frac{1}{b}$ and the central charge $c=1+6Q^{2}$, up to a $c$-number anomaly \cite{Zamolodchikov:1995aa}.

The Euclidean space-like Liouville partition function, with a canonically normalized Liouville field $\phi$, can be written as
   \be\label{spacelikeaction}
  \mathcal{Z} =  \int\D\phi \exp{\Big[-\frac{1}{4\pi}\int dx^2 \sqrt{\tilde{g}}\big(\tilde{g}^{ab}\partial_{a}\phi\partial_{b}\phi +Q\tilde{R}\phi + 4\pi\mu e^{2b\phi}\big)\Big]}.
   \ee
   
 This form depends on the fact that the metric can be gauge fixed in a generally covariant way to conformal gauge i.e. the \emph{Physical} metric, $g_{\mu\nu}$  can be written in terms of the product of the exponentiated Liouville factor and a \emph{Reference} metric $\tilde{g}_{\mu\nu}$ giving $g_{\mu\nu} = e^{2b\phi}\tilde{g}_{\mu\nu}$\cite{Zamolodchikov:2005fy}. To make contact with the classical Liouville equation, the $1/b^{2}$ dependence  of the central charge
which has been absorbed into the definition of the Liouville field must be taken into account in order to canonically normalize the action. In the Semi-classical limit, the action can be written in terms of classical field via the field redefinition $\phi_{c} = 2b\phi$,
\be\label{spacelikeactionclassical}
-\frac{1}{16\pi b^{2}}\int dx^2 \sqrt{\tilde{g}}\big(\tilde{g}^{ab}\partial_{a}\phi_{c}\partial_{b}\phi_{c} + 2bQ\tilde{R}\phi_{c} + 16\pi\mu b^{2} e^{\phi_{c}}\big).
\ee

Here the dominant contribution of the central charge $c\propto 1/b^{2}$ has been factored out. To make a good semi-classical limit the ``cosmological constant" $\mu$ must scale as
$1/b^{2}$. The actual cosmological constant $\lambda =\pi\mu b^{2}$, is well defined in the semi-classical $b\rightarrow 0$ limit.\footnote{The value $\lambda$ is a tunable constant in the Liouville theory. It can be changed by adding a constant linear shift to the Liouville field. The value of $\lambda$ will be set by the radius of the sphere in what follows.} Timelike Liouville results from
(\ref{spacelikeaction}) under the continuation $b \rightarrow -i\hat{b}$, $\phi \rightarrow i\hat{\phi}$ and $Q\rightarrow i\hat{Q}$ with $\hat{b} \in \mathbb{R}$. The resulting action is
    \be\label{timelikeaction}
-\frac{1}{4\pi}\int dx^2 \sqrt{\tilde{g}}\big(-\tilde{g}^{ab}\partial_{a}\hat{\phi}\partial_{b}\hat{\phi} - \hat{Q}\tilde{R}\hat{\phi} + \frac{4\lambda}{\b^{2}}
e^{2\hat{b}\hat{\phi}}\big).
   \ee
 In the semi-classical limit, $\hat{b} \rightarrow 0$, (\ref{timelikeaction}) has a large negative central charge $c = 1 -6\hat{Q}^2$ with $\hat{Q} =1/\hat{b}-\hat{b}$.
   
One cannot simply compute the partition function for  (\ref{timelikeaction}) by simply integrating  (\ref{timelikeaction}) over all fluctuations about the sphere, since the kinetic term in (\ref{timelikeaction}) is the wrong sign and the path integral is formally divergent. One must take the path integral of the partition function of (\ref{spacelikeaction}) and analytically continue it, taking Stokes Phenomenon into account employing the results of \cite{Harlow:2011ny} to define the Timelike partition function. Since all the relevant saddles of the integration cycle, not just the sphere, must be taken into account to get finite answers and reproduce exact results like the Timelike DOZZ formula\footnote{This continuation property is what allows us to use a path integral approach to compute Timelike Liouville correlation functions. As the wrong sign kinetic term of (\ref{timelikeaction}) renders the partition function integral formally divergent if Stokes Phenomenon isn't taken into account\cite{Harlow:2011ny}. In this paper it will be assumed that this has already been taken into account. A complete account of Stokes Phenomenon and the Timelike partition function goes beyond the scope of this paper, for a nice account one should look at
\cite{Harlow:2011ny,Witten:2010cx,Bender:1978:AMM}.}.
   
The action of Timelike Liouville has the 2 sphere, $\mathbb{S}^2$ as homogeneous real saddle point. \footnote{Viewed from the Space-like side this is a complex saddle point} The saddle point of the field is defined by the
constant Liouville field value
  \be\label{constsaddle}
  \hat{\phi} = \hat{\phi}_0 =\frac{1}{2\b}\log{\Big|\frac{\hat{Q}\b\tilde{R}}{8\lambda}\Big|}.
  \ee
  
Perturbations by ``\emph{light}" operators\footnote{The terminology ``\emph{Light}" and ``\emph{Heavy"} primary operators,
is standard in the study of Liouville theory. When computing correlators of primary operators $<e^{\alpha_1\phi_1}\ldots e^{\alpha_n\phi_n}>$, an operator is called \emph{heavy} if its
Liouville momentum $\alpha_i \sim \frac{\sigma_i}{\b}$ in the $\b\rightarrow 0$ limit, and \emph{light} if $\alpha_i \sim \b\sigma_i$ as $\b\rightarrow 0$. Heavy operators can effect the
classical saddle point, as they scale in the same way as the action while light operators give sub-leading contributions. }, which scale as $\b\sigma$ in Liouville momentum will not effect the saddle point and hence a perturbative expansion of
(\ref{timelikeaction}) about the spherical saddle point can be made without changing the saddle point, i.e. fluctuations cannot change the topology. Expanding the Liouville field as $\hat{\phi} = \hat{\phi}_0 + f$ and
expanding to quadratic order in $\b$, yields a quadratic action for $f$, which apart from an irrelevant constant $S_0$, is independent of the value of $\lambda$.
  
  \be\label{linearactionint}
-\frac{1}{4\pi}\int dx^2 \sqrt{\tilde{g}}\big(-\tilde{g}^{ab}\partial_{a}f\partial_{b}f +\tilde{R}f^2\big) + S_0.
    \footnote{From the Liouville saddle point (\ref{constsaddle}) and what later follows in Section \ref{dist} this will imply that $\lambda = 1/4$.  However to aid in the analysis $\lambda$ will be left general for now and determined later.}\ee  Expressing the action (\ref{linearactionint}) in spherical coordinates and integrating by parts yields
  \be\label{laction1}
-\frac{1}{4\pi}\int d\theta d\varphi\sin{\theta}\big\{f\Big(\frac{1}{\sin{\theta}}\partial_{\theta}\sin{\theta}\partial_{\theta} +\frac{1}{\sin^2{\theta}}\partial^2_{\varphi}
+ 2\Big)f\big\}. \footnote{Note $R= 2$ for the unit sphere.}
    \ee
\begin{figure}[ht]\label{dipole}
\begin{center}
\includegraphics[scale=0.6]{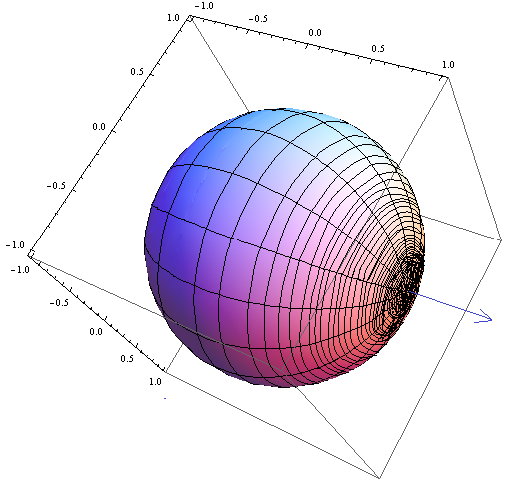}
\caption{A sphere embedded in $\mathbb{R}^{3}$  with a coordinate system possessing a dipole. There is more coordinate area on one side than the other. The blue arrow denoting the dipole vector.}
 \label{dipole}
\end{center}
\end{figure}

This action has a zero mode which must be  dealt with in order to compute quantities in perturbation theory. The zero mode corresponds to the $SL_{2}(\mathbb{C})$ conformal coordinate transformations that can be
performed on the coordinates of the reference sphere. These transformations have the effect of moving coordinate area around the sphere. The non-compact portion of this gauge redundancy can be attributed to
the overall dipole of area that the physical manifold has compared to the reference sphere; see figure \ref{dipole}. This last gauge freedom must be dealt with using a Fadeev-Popov procedure.
  
The equation of motion resulting from (\ref{laction1}) is that massive scalar field on a sphere. The calculation can be simplified by exploiting the fact that the Green's function will only depend on the geodesic separation between points on the
sphere has only one singularity and  is rotationally symmetric around that singularity\footnote{The non trivial fact that a massive scalar field on a sphere can possess a single singularity unlike the massless case which must have two, is due to mass causing field lines to be die off before they reach the other side of the sphere to form a second singularity}. This rotational symmetry implies that the Green's function will only depend on the angle between the source point and the field point, $\theta, \varphi$ and $\theta^\prime,\varphi^\prime$. This
means that the Green's function $G$ is only a function of $\cos{\beta} = \vec{x}\cdot\vec{x}^\prime = \cos{\theta}\cos{\theta^\prime} +\sin{\theta}\sin{\theta^\prime}\cos{(\varphi - \varphi^\prime)}$. Using the rotational symmetry of the differential operator and calling $\chi = \cos{\beta}$ we can rewrite the Green's function equation from (\ref{laction1}) into
  
  \be\label{laction2}
\big(\partial_{\chi}(1 - \chi^2)\partial_{\chi} + 2\big)G = \frac{1}{2\pi}\delta{[1-\chi]}.
  \ee

\begin{figure}[ht]\label{dipole2}
\begin{center}
\includegraphics[scale=0.6]{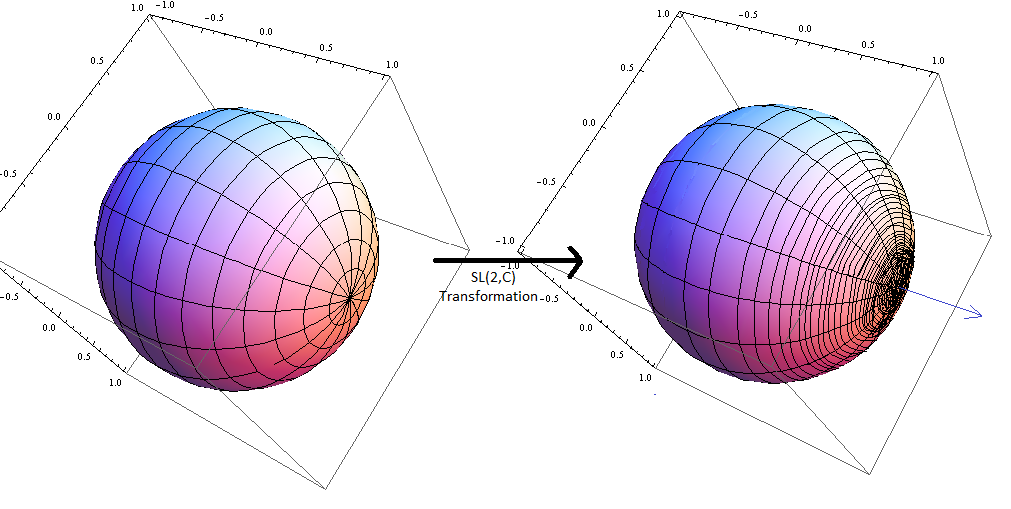}
\caption{An SL(S,C) transformation, changes the dipole of the coordinate system of the sphere}
 \label{dipole2}
\end{center}
\end{figure}
\hspace{0.25in}Eq. (\ref{laction2}) still possesses a zero mode. The zero mode is due to invariance of the physical geometry under $SL_{2}(\C)$ reparametrizations of the coordinates on the reference sphere which results in a compensating change of the Liouville field, leaving the physical metric invariant.  This $SL_{2}(\mathbb{C})$ transformation is a combination of a compact rotations of the coordinate system and/or a non-compact change in the dipole of coordinate system. This can be visualized as follows, if the coordinates are thought of as a charge density on a sphere embedded in $\mathbb{R}^{3}$, $SL_(2,\mathbb{C})$ transformations change the dipole of this charge density by pushing points toward one site on the sphere and repelling them from another. This increases the local charge density about one site of the sphere and decreases about another. In terms of the coordinates it increases the amount of coordinate area about one point and decreases it about another, see Figure \ref{dipole2}. If the dipole of the Liouville field is fixed as a gauge condition and the Faddeev-Popov procedure of integrating over the gauge condition is implemented, a Green's function can be obtained for use in perturbation theory. In (\ref{laction2}) the statement that there is a zero mode remaining is just the fact that a solution of (\ref{laction2}) is invariant under $G \rightarrow G+\alpha_1 P_1(\chi)= G + \alpha_1\chi $ for any value of $\alpha_1$.\footnote{There is only one $\alpha_1$ here because the coordinates have been rotated into the direction of the dipole, and $\alpha_1$ is the magnitude of the dipole vector. If this were not exploited, then $\alpha_1$ would be replaced by the vector $\vec{\alpha}$ and the zero mode statement would be $G \rightarrow G+\sum^{1}_{m=-1}\alpha_m Y^{(1,m)}[\theta,\varphi].$}
 In these rotated coordinates (\ref{laction2}) is just the equation for the Green's function for Legendre's differential equation with $l = 1$.
  
  To proceed, the Fadeev-Popov procedure is employed\cite{Faddeev196729,Faddeev:1969su} on (\ref{laction1}) with the gauge condition that the dipole over the sphere is fixed and then integrated over all values. The dipole is defined as,
  \be\label{gaugecondition}
  \vec{n} = \int d\theta\,d\varphi\sin{\theta}\Big(G\vec{x}[\theta,\varphi]\Big) .\footnote{Here $\vec{x} = |x|(\sin{\theta}\cos{\varphi}\hat{\mathbb{I}} +\sin{\theta}\sin{\varphi}\hat{\mathbb{J}}+\cos{\theta}\hat{\mathbb{K}})$ is the $\mathbb{R}^{3}$ position vector in spherical coordinates. It is imaged that the reference sphere is embedded in a larger $\mathbb{R}^{3}$.}
  \ee
  The functional delta function is inserted into the partition function for (\ref{laction1}) by inserting the identity 
  
  \be\label{fadeevpopov}
  1 = \int d^{3}\alpha\,\Delta_{FP}\,\delta^{(3)}
  \Big[ \int d\theta\,d\varphi\sin{\theta}\Big(G^{(\vec{\alpha})}\vec{x}[\theta,\varphi]\Big)- \vec{\kappa}\Big]
  \ee
into the functional integral for the partition function.
  Here $G^{(\vec{\alpha})}$ means the gauge transformed $G$ with parameters $\vec{\alpha}$, and $\Delta_{FP}$ is the Fadeev-Popov determinant.
  Following the standard methods of inserting (\ref{fadeevpopov}) into partition function with action (\ref{laction1}), transforms the linearized Timelike Liouville partition function into
  
  \begin{align}\label{funtionalintegral}
&\mathcal{N}\int d^{3}\kappa e^{ -\frac{\Lambda}{2}\vec{\kappa}\cdot\vec{\kappa}}\int d^{3}\alpha\int\D f \\&\hspace{1in}\times\exp{\Big[-\frac{1}{4\pi}\int d\theta\,d\varphi\sin{\theta}\big\{f\Big(\frac{1}{\sin{\theta}}\partial_{\theta}\sin{\theta}\partial_{\theta} +\frac{1}{\sin^2{\theta}}\partial^2_{\varphi} + 2\Big)f\big\}\Big]}\nonumber\\&\hspace{2in}\times\delta^{(3)}(\big\{\int d\theta\,d\varphi\sin{[\theta]}\vec{x} f^{(\alpha)}) \big\}- \vec{\kappa})\Delta_{FP}\nonumber.
  \end{align}
  
  Integrating over $\vec{\kappa}$ results in 
  \begin{align}\label{fgaugedint}
&\mathcal{N}\Delta_{FP}\int d^{3}\alpha\int\D f \\&\hspace{0.45in}\times\exp{\Big[-\frac{1}{4\pi}\int d\theta\,d\varphi\sin{\theta}\Big\{\Big\{f\Big(\frac{1}{\sin{\theta}}\partial_{\theta}\sin{\theta}\partial_{\theta} +\frac{1}{\sin^2{\theta}}\partial^2_{\varphi} + 2\Big)f\Big\} -2\pi\Lambda f\vec{x}\cdot\vec{n}\Big\}\Big]}.\nonumber
  \end{align}
 From the point of view of the Green's function equation obtained from (\ref{fgaugedint}), the integral (\ref{gaugecondition})  is just a c-number vector. Hence the Green function equation resulting from (\ref{fgaugedint}) can be solved by treating (\ref{gaugecondition})  as a constraint on the Green's function and then enforcing that constraint to obtain a final answer. The Green's function equation resulting from (\ref{fgaugedint}) taking into account the fact that the Green's function only depends on $\chi$ is
  \be\label{greenfunctioneqn}
  \partial_{\chi}\{(1 - \chi^2)\partial_{\chi}G\} + 2G = -\pi\Lambda n\chi + \frac{1}{2\pi}\delta(1 - \chi),
  \ee
with
\be\label{gaugecon}
n = 2\pi\int^{1}_{-1}d\chi^{\prime} \,\chi^{\prime}G[\chi^\prime].
\ee

The gauge fixing procedure has resulted in a inhomogeneous term in (\ref{greenfunctioneqn}), this term can be interpreted as a background charge that absorbs the field lines coming from
the singularity at $\chi = 1$ so that the field is smooth everywhere else on the sphere.
  The general solution to (\ref{greenfunctioneqn}) when $\chi \neq 1$  is
  \be
G = \alpha_1\chi + \Big(\frac{\pi n\Lambda}{6} + \frac{c_2}{2}\Big)\chi\log{|1 + \chi|} + \Big(\frac{\pi n\Lambda}{6} - \frac{c_2}{2}\Big)\chi\log{|1 - \chi|} - c_2.
  \ee
  
Imposing the boundary conditions of finiteness $\chi \neq 1$, smoothness of the solution at $\chi = -1$, normalizing the Green's function so (\ref{greenfunctioneqn}) is obeyed  when $\chi = 1$, and imposing the
constraint equation (\ref{gaugecon}) gives values for $\Lambda$, $\alpha_1$, $n$, $c_2$. 

The resulting Green's function is
  
  \begin{align}\label{greenfunction}
 \text{\begin{fmffile}{liouvprop}
\begin{fmfgraph*}(25,3)
\fmfleft{in}
\fmfright{out}
\fmflabel{$x$}{in}
\fmflabel{$y$}{out}
\fmf{dashes}{in,out}
\fmfdot{in,out}
\end{fmfgraph*}
\end{fmffile}}
 \hspace{6pt}&=<f[\theta_x,\varphi_x]f[\theta_y,\varphi_y]>\\&= \frac{1}{8\pi^2}\Big\{-(\log{2} + 1/2)\chi_{xy}+ \chi_{xy}\log{|1 - \chi_{xy}|} + 1\Big\}\nonumber
  \end{align}
  
with $\Lambda = -\frac{9\cdot3}{11\cdot2\pi^2} $, $n = -\frac{11}{4\cdot9\pi}$, $\alpha_1 =-\frac{1}{8\pi^2}(\log{2} + 1/2)$, and $c_2 = -\frac{1}{8\pi^2}$; and $\chi_{xy}
=\cos{\theta_x}\cos{\theta_y} +\sin{\theta_x}\sin{\theta_y}\cos{[\varphi_x - \varphi_y]}$ .

Quantities can now be computed in perturbation theory using Wicks theorem. Which relates the Green's function to the two point correlator of the field\cite{PhysRev.80.268}.  
  This is the Green's function that will be used to compute quantities in perturbation theory.
 \section{Perturbative Correction To The Geodesic Distance Between Two Points On The Bumpy Sphere.}\label{dist}
\hspace{0.25in}
Now that the zero mode has been dealt with, the perturbative correction of the expectation value of physical geodesic distance between two arbitrary points lying on a north south trajectory on the reference sphere can now be computed.\footnote{  This is more general then it seems, since one can simply rotate the coordinates of the reference sphere to move two arbitrary points onto a north-south trajectory, no generality is lost by computing north-south distances.}  The advantage of computing north/south trajectories is that the two end points will have the same value of the azimuthal angle $\varphi$ which will be  called $\varphi_0$ and this simplifies the calculation\footnote{Because $\phi$ is the standard symbol for the Liouville field, $\varphi$ will be used for the azimuthal
angle.}.

The quantity that will be studied is the expectation value of geodesic distance $L$ between two points on a north south trajectory computed up to second order in $\b$\footnote{Here the reference polar angle $\theta$ has been chosen as the parameter along the geodesic, to avoid any vielbien ambiguities.}

\begin{align}\label{geodesicdist}
L&= \Big<\int^{\theta_2}_{\theta_1}\sqrt{g_{\mu\nu}\frac{\partial x^{\mu}}{\partial\theta}\frac{\partial x^{\nu}}{\partial\theta}}\,d\theta\Big>\\&= \Big<\int^{\theta_2}_{\theta_1}\sqrt{e^{2\b\phi_{0}+2\b f}\tilde{g}_{\mu\nu}\Big\{1 + \sin^{2}{\theta}\Big(\frac{d\varphi}{\partial\theta}\Big)^{2}\Big\}}\,d\theta\Big>\nonumber\\
&=\frac{\mathcal{N}\Delta_{FP}\int d\alpha\mathcal{D} f \Big(\int^{\theta_2}_{\theta_1}\sqrt{e^{2\hat{b}\phi_0 +2\hat{b} f}\tilde{g}_{\mu\nu}\frac{\partial x^{\mu}}{\partial\theta}\frac{\partial x^{\nu}}{\partial\theta}}\,d\theta\Big)e^{-S_{\text{gauge fixed action}}}}{\mathcal{N}\Delta_{FP}\int d\alpha\mathcal{D} f e^{-S_{\text{gauge fixed action}}}}.\nonumber
\end{align}
 The unperturbed geodesic is the portion of the latitude line connecting $\theta_1$ and $\theta_2$, implying that the unperturbed geodesic $\varphi_0$ is a constant.
Variation with respect to $x^{(2)}=\varphi$ in (\ref{geodesicdist}) results in the usual geodesic equation,
\be\label{Lvar1}
\delta L = -\int^{\theta_2}_{\theta_1}d\theta\,\Big(\frac{d^{2}\varphi}{d\theta^{2}} +\Gamma^{\varphi}_{\mu\nu}\frac{dx^{\mu}}{d\theta}\frac{dx^{\nu}}{d\theta} \Big) g_{\varphi\varphi}\delta\varphi = 0.
\ee It follows that, corrections from the reference geodesic equation comes from two sources:
the explicit factor of $e^{2\b\phi}$ in (\ref{geodesicdist}), and the change in the Christoffel symbol that results from it,
\be\label{Christoffel}
\Gamma^{\lambda}_{\mu\nu} = \tilde{\Gamma}^{\lambda}_{\mu\nu} + b\tilde{g}^{\lambda\sigma}\big[(\partial_{\mu}f)\tilde{g}_{\nu\sigma} + (\partial_{\nu}f)\tilde{g}_{\sigma\mu} -
(\partial_{\sigma}f)\tilde{g}_{\mu\nu}\big].
\ee

Here $f$ is the fluctuation in the Liouville field from $\hat{\phi}_{0}$. The geodesic equations derived from (\ref{Christoffel}) are\footnote{
$\hat{\Gamma}^{\varphi}_{\varphi\theta} = \cot{\theta}$ is the only pertinent non-zero Christoffel symbol for the reference geometry of $\S^{2}$.}
\be\label{geodeqn}
\frac{d^{2}\varphi}{d\theta^{2}} + 2\tilde{\Gamma}^{\varphi}_{\varphi\theta}\Big(\frac{d\varphi}{d\theta}\Big)
=-\b\Big[2\frac{d\varphi}{d\theta}\Big(\partial_{\theta}f + \partial_{\varphi}f\Big) - 
\frac{\partial_{\varphi}f}{\sin^{2}{\theta}}\Big\{1 + \sin^{2}{\theta}\Big(\frac{d\varphi}{d\theta}\Big)^{2}\Big\}\Big].
\ee
Inserting (\ref{geodeqn}) into (\ref{Lvar1}) gives
\begin{align}\label{Lvar2}
\delta L&=-\int^{\theta_2}_{\theta_1}d\theta\,\sin^{2}{\theta}e^{2\b f}\delta\varphi\Big\{\frac{d^{2}\varphi}{d\theta^{2}} +2\cot{\theta}\frac{d\varphi}{d\theta} + \b\Big[2\frac{d\varphi}{d\theta}\Big(\partial_{\theta}f + \partial_{\varphi}f\Big) \\
&\hspace{3in}-\frac{\partial_{\varphi}f}{\sin^{2}{\theta}}\Big(1 + \sin^{2}{\theta}\Big(\frac{d\varphi}{d\theta}\Big)^{2}\Big)\Big]\Big\}\nonumber\\
&=-\int^{\theta_2}_{\theta_1}d\theta\,e^{2\b f}\delta\varphi\Big\{\frac{d}{d\theta}\Big(\sin^{2}{\theta}\frac{d\varphi}{d\theta}\Big) - \b
\partial_{\varphi}f \nonumber\\&\hspace{2in}+ \b\sin^{2}{\theta}\Big[2\frac{d\varphi}{d\theta}\partial_{\varphi}f +\Big(\frac{d\varphi}{d\theta}\Big)^{2} \partial_{\varphi}f \Big]\Big\}.\label{Lvar3}
\end{align}

Expressing the corrections in the geodesic  $\varphi[\theta]$ as,

\be\label{gcorr}
\varphi[\theta] = \varphi_0 +\b\varphi_{1}[\theta] + \b^{2}\varphi_{2}[\theta] + \ldots
\ee
and substituting (\ref{gcorr}) into (\ref{Lvar3})  leads to a set of equations of different orders in $\b$. The zeroth, first, and second order equations are respectfully,

\begin{align}
\frac{d}{d\theta}\Big(\sin^{2}{\theta}\frac{d\varphi_0}{d\theta}\Big)&=0\label{firstord}\\
\vspace{0.25in}\nonumber\\
\frac{d}{d\theta}\Big(\sin^{2}{\theta}\frac{d\varphi_1}{d\theta^{2}}\Big)=\partial_{\varphi}f-\sin^{2}{\theta}&\Big[2\frac{d\varphi_0}{d\theta}\partial_{\varphi}f + \Big(\frac{d\varphi_0}{d\theta}\Big)^{2} \partial_{\varphi}f \Big]\label{secondord}\\
\vspace{0.25in}\nonumber\\
\frac{d}{d\theta}\Big(\sin^{2}{\theta}\frac{d\varphi_2}{d\theta^{2}}\Big)=\sin^{2}{\theta}\Big[2\frac{d\varphi_1}{d\theta}&\partial_{\varphi}f + 2\frac{d\varphi_0}{d\theta} \frac{d\varphi_1}{d\theta} \partial_{\varphi}f \Big]\label{thirdord}.
\end{align}
\\
As was mentioned previously, $\varphi_0$ is a constant. This is consistent with (\ref{firstord}) and also implies that (\ref{secondord}) reduces to
\be\label{geodesicequ1}
\frac{d}{d\theta}\Big(\sin^{2}{\theta}\frac{d\varphi_1}{d\theta}\Big)=\partial_{\varphi}f.
\ee
Lastly, since $L$ is being computed to second order in $\b$, (\ref{geodesicdist}) implies that only the first order correction $\varphi_1$ is needed, and (\ref{thirdord}) is not necessary.

The classical action that generates (\ref{geodesicequ1}) up to total derivatives, is that of a forced harmonic oscillator with vanishing kinetic term\footnote{A classical action that generates (\ref{geodesicequ1}) up to total derivatives is \be\nonumber
\int^{\theta_2}_{\theta_1}\Big[\frac{1}{2}\sin^{2}{\theta}\Big(\frac{d\varphi_1}{d\theta^{2}}\Big)^{2} + f\Big]d\theta = \int^{\theta_2}_{\theta_1}\Big[\frac{1}{2}\Big\{\Big(\frac{du_1}{d\theta}\Big)^{2} - u^{2}_1\Big\} + f\Big]d\theta
\ee.
The time evolution parameter in this case be $\theta$.}.
 A good conjugate variable to describe the system is then $u = \varphi\sin{\theta}$, which rewrites (\ref{gcorr}) as
\begin{align}\label{gcorru}
u &= u_0 + \b u_1[\theta] + \b^{2}u_2[\theta] + \ldots\\ &= \varphi[\theta]\sin{\theta}  \nonumber\\&= \varphi_0 \sin{\theta} +\b\varphi_{1}[\theta]\sin{\theta} + \b^{2}\varphi_{2}[\theta] +\ldots\nonumber,
\end{align} changes (\ref{firstord}) to
\be\label{geodeqn1}
\frac{d}{d\theta}\Big(\sin{\theta}\Big[\frac{du_0}{d\theta} - \frac{u_0\cos{\theta}}{\sin{\theta}}\Big]\Big)= \sin{\theta}\Big(\frac{d^{2}u_0}{d\theta^{2}}  +u_0\Big) = 0
\ee
and changes (\ref{geodesicequ1}) into 
\be\label{geodeqn2}
\frac{d^{2}u_1}{d\theta^{2}} + u_1 =\partial_{u}f.
\ee

This rewriting makes the following computations easier.
Since the geodesics begin and end on the same value of $\varphi = \varphi_0$; our boundary conditions are that $u_{0}[\theta_2] = \varphi_0\sin{\theta_2}$, $u_{0}[\theta_1] = \varphi_0\sin{\theta_1}$, and that $u_{1}[\theta_2] = u_{1}[\theta_1] = 0$.\footnote{These boundary conditions are the correct ones as the unperturbed geodesic was just the path $\varphi[\theta]=\varphi_0$ on the reference sphere. Small fluctuations of the sphere, ``bumps", do not change the $\varphi$ position of points hence the end points do not move; therefore the correction $u_1$ should vanish at at the end points.}
These equations can formally be solved to create an expansion for $u$ up to order $\b$,
\begin{align}\label{ueqn}
u = \varphi_0\sin{\theta}\, + \, \b\Big\{\int^{\theta}_{\theta_1}d\hat{\theta}&\sin{[\theta - \hat{\theta}]}\partial_{u}f[\hat{\theta}]\nonumber\\& - \frac{\sin{[\theta_2 -\hat{\theta}]}}{\sin{[\theta_2 - \theta_1]}}\int^{\theta_2}_{\theta_1}d\hat{\theta}\sin{[\theta_2 -\hat{\theta}]}\partial_{u}f[\hat{\theta}]\Big\}.
\end{align}

Rewriting the expectation value (\ref{geodesicdist}) in terms of the variable $u$ yields

\begin{align}\label{mainquant}
L&=\Bigg< e^{b\phi_0}\int^{\theta_2}_{\theta_1}d\theta e^{\b f}\sqrt{1+\Big\{\frac{du}{d\theta} - \frac{u\cos{\theta}}{\sin{\theta}}\Big\}^{2}}\Bigg> \\ \nonumber\text{with}\hspace{1.25in}&\nonumber\\
\Big\{\frac{du}{d\theta} - \frac{u\cos{\theta}}{\sin{\theta}}\Big\}^{2}& = \frac{\b^{2}}{\sin^{2}{\theta}}\Bigg\{\frac{\sin{\theta_1}}{\sin{[\theta_2 - \theta_1]}}\int^{\theta_2}_{\theta_1}d\hat{\theta}\sin{[\theta_2 - \hat{\theta}]}\partial_{u}f[\hat{\theta}]\\&\hspace{2.5in}-\int^{\theta}_{\theta_1}d\hat{\theta}\sin{\hat{\theta}}\partial_{u}f[\hat{\theta}]\Bigg\}^{2}\nonumber.
\end{align}

The second line results from (\ref{ueqn})\footnote{Since (\ref{mainquant}) is being expanded to order $\b^{2}$, it is valid to evaluate $u$ only to first order in $\b$ as the lowest order correction under the square root sign is $\b^{2}$, (the zeroth order term drops out). Higher order corrections of $u$ will only contribute O($\b^3$) corrections.}. It is evident from (\ref{mainquant}) that $e^{b\phi_{0}} =\Big(\frac{\tilde{R}}{8\lambda}(1 - \b^{2})\Big)^{1/2}$ is the radius of the sphere which has been set to $1$ in the $\b\rightarrow0$ limit. Recalling that $\tilde{R} = 2$, this shows that the value of $\lambda = 1/4$ if $\theta_2-\theta_1$ is to be interpreted as the difference in polar angle for the unperturbed path.\footnote{ This makes sense as $\lambda^{-1/2}$ has units of radius of curvature as can be seen from  the classical equation of motion from (\ref{timelikeaction}) in the semi-classical limit. Specifically $R \propto \lambda$ \cite{Ginsparg:1993is}. Since constant shifts $\phi$ can be used to tune the value of $\lambda$,  it  sets the radius of the sphere\cite{Nakayama:2004vk}.} Expanding (\ref{mainquant}) up to  and including $\mathcal{O}[\b^{2}]$ results in\footnote{The one point function $\big<f\big> =0$ by the symmetry of the linearized action(\ref{linearactionint}). Since (\ref{linearactionint}) is quadratic in $f$ having a non zero value of \big<f\big> means the field is not fluctuating about its minimum.}

\begin{align}\label{mainquantpert}
L =\Big(\frac{1}{4\lambda}\Big)^{1/2}\Big\{&\int^{\theta_2}_{\theta_1}d\theta\Big(1  -  \frac{\b^{2}}{2}  +\frac{1}{2}\b^{2}\big<f[\theta]f[\theta]\big> \\&+\frac{\b^{2}}{2\sin^{2}{\theta}}\Big\{\frac{\sin^{2}{\theta_1}}{\sin^{2}{[\theta_2 - \theta_1]}}\int^{\theta_2}_{\theta_1}\int^{\theta_2}_{\theta_1}d\hat{\theta}\,d\bar{\theta}\sin{[\theta_2 - \hat{\theta}]}\sin{[\theta_2 - \bar{\theta}]} \nonumber\\&\hspace{0.5in} - 2\frac{\sin{\theta_1}}{\sin{[\theta_2 - \theta_1]}}\int^{\theta}_{\theta_1}\int^{\theta_2}_{\theta_1}d\hat{\theta}\,d\bar{\theta}\sin{\hat{\theta}}\sin{[\theta_2-\bar{\theta}]}\nonumber\\ &\hspace{1in}+\int^{\theta}_{\theta_1}\int^{\theta}_{\theta_1}d\hat{\theta}\,d\bar{\theta}\sin{\hat{\theta}}\sin{\bar{\theta}}\sin{\hat{\theta}}\Big\}\big<\partial_{u}f[\hat{\theta}]\partial_{u}f[\bar{\theta}]\big>\Big\}\nonumber.
\end{align}
Looking at (\ref{greenfunction})  evaluated when $\varphi_x = \varphi_y$, it is evident that the correlator made out of descendants $\big<\partial_{u}f[\theta]\partial_{u}f[\theta^\prime]\big>$ can be obtained, by taking the appropriate derivatives of (\ref{greenfunction}) and then setting $\varphi_x = \varphi_y =\varphi_0$.

\begin{align}\label{descendantcorr}
<\partial_{u}f[\hat{\theta}]\partial_{u}f[\bar{\theta}]>&= \frac{1}{8\pi^2}\Big\{-(\log{2} + 1/2)+ \log{|1 - \cos{[\hat{\theta} - \bar{\theta}]}|} \\&\hspace{1in}- \frac{\cos{[\hat{\theta} - \bar{\theta}]}}{1 - \cos{[\hat{\theta} - \bar{\theta}]}}\Big\}\nonumber.
\end{align}

There are two issues in proceeding further; first both $\big<\partial_{u}f[\theta]\partial_{u}f[\theta^\prime]\big>$ and $\big<f[\theta]f[\theta^\prime]\big>$ diverge as $\theta\rightarrow\theta^\prime$, and second, there are three non-trivial integrals that involve $\big<\partial_{u}f[\theta]\partial_{u}f[\theta^\prime]\big>$. The coincident divergence problem is treated by introducing a short distance regulator $\epsilon$, into both $\big<f[\theta]f[\theta^\prime + \epsilon]\big>$ and $\big<\partial_{u}f[\theta]\partial_{u}f[\theta^\prime+\epsilon]\big>$, evaluating (\ref{mainquantpert}) with the regulator in place, and finally taking $\epsilon\rightarrow0$\footnote{It should be noted that the $\big<f[\theta]f[\theta]\big>$ term in (\ref{mainquantpert}) is manifestly divergent. This is due to the fact that both variables are evaluated at the same point. This divergence is logarithmic, as can be seen when the regulator $\epsilon$ is added, $\big<f[\theta]f[\theta+\epsilon]\big> = \frac{1}{8\pi^2}\big(-(\log{2} + 1/2)\cos{\epsilon}+ \cos{\epsilon}\log{|1 - \cos{\epsilon}|} + 1\big)$. This $\log$, is the factor that is cancelled by the integrals involving $\big<\partial_{u}f[\theta]\partial_{u}f[\theta^\prime+\epsilon]\big>$ in (\ref{mainquantpert}). All other divergent quantities in $\big<\partial_{u}f[\theta]\partial_{u}f[\theta^\prime+\epsilon]\big>$ integral cancel internally, leaving a finite result.}. The second problem is more technical, brute force calculation of (\ref{mainquantpert}) results in a proliferation of terms to be computed. The calculation is simplified dramatically if the following trick is employed; rewriting the correlator as follows
\begin{align}\label{trick}
\big<\partial_{u}f[\hat{\theta}]\partial_{u}f[\bar{\theta} +\epsilon]\big> =\int^{\pi}_{0}\int^{\pi}_{0}d\alpha\,d\beta\,\delta[\alpha - \hat{\theta}]\delta[\beta - \bar{\theta}]\big<\partial_{u}f[\alpha]\partial_{u}f[\beta +\epsilon]\big>,
\end{align}
 and placing this into (\ref{mainquantpert}) allows all the pre-factors of  $\big<\partial_{u}f[\hat{\theta}]\partial_{u}f[\bar{\theta} +\epsilon]\big>$ to be integrated. (\ref{mainquantpert}) is reduced to

\begin{align}\label{mainquantpertint}
L =\Big(\frac{1}{4\lambda}\Big)^{1/2}\Big\{&(\theta_2 - \theta_1)\Big(1  -  \frac{\b^{2}}{2} +\frac{1}{2}\b^{2}\big<f[\theta]f[\theta+\epsilon]\big>\Big) \\&+\frac{\b^{2}}{2}\Big\{\int^{\theta_2}_{\theta_1}\int^{\theta_2}_{\theta_1}d\alpha\,d\beta\frac{\sin{[\theta_2 - \beta]}\sin{[\alpha - \theta_1]}}{\sin{[\theta_2 - \theta_1]}} \nonumber\\&\hspace{0.5in}- \int^{\theta_2}_{\theta_1}\int^{\alpha}_{\theta_1}d\alpha\,d\beta\sin{[\alpha - \beta]}\Big\}\big<\partial_{u}f[\hat{\theta}]\partial_{u}f[\bar{\theta} + \epsilon]\big>\Big\}\nonumber.
\end{align}

In the $\b\rightarrow0$, $L$ should reduce to the geodesic length on the unperturbed unit sphere implying $\lambda = 1/4$. These last integrals can now be evaluated with less but still considerable effort. Once (\ref{mainquantpertint})  is evaluated at finite $\epsilon$, the Log divergence from $\big<f[\theta]f[\theta+\epsilon]\big>$ cancels the remaining Log divergence from the $\big<\partial_{u}f[\hat{\theta}]\partial_{u}f[\bar{\theta} +\epsilon]\big>$ integrals. Apart from the one Log divergence that cancels the $\big<f[\theta]f[\theta+\epsilon]\big>$ divergence, all other factors of $\log{[1-\cos{\epsilon}]}$ originating from the  $\big<\partial_{u}f[\hat{\theta}]\partial_{u}f[\bar{\theta} +\epsilon]\big>$ integrals cancel amongst themselves  in the limit $\epsilon \rightarrow0$ and the result of (\ref{mainquantpert}) is finite
\begin{align}\label{result1}
L = \,&(\theta_2 - \theta_1)(1 - \b^{2}/2) +   \frac{\b^{2}}{16\pi^{2}}\Big\{-(\log{2} + 1/2)\big\{\sin{[\theta_2 - \theta_1]}\\& + (1 - \cos{[\theta_2 - \theta_1]})\tan{\Big[\frac{\theta_2 - \theta_1}{2}\Big]}\big\} -2(\theta_2 - \theta_1) \nonumber\\&\hspace{0.5in}+ 2\tan{\Big[\frac{\theta_2 - \theta_1}{2}\Big]}\log{|1 - \cos{[\theta_2 - \theta_1]}|}\Big\}\nonumber.
\end{align}
\section{Results And Discussion}\label{res}

\subsection{Interpretation Of The Finiteness Of \texorpdfstring{$L$}{L} To Second Order In \texorpdfstring{$\b$}{b}.}

\hspace{0.25in}The order $\b^{2}$ correction has two contributions. The contribution proportional to $\frac{\b^{2}}{16\pi^{2}}$, is the main result of the perturbative computation.  One possible surprising result is that $L$ is finite at all for non-zero separation angle. One possible intuition due to power counting is that in higher dimensions that $L$ would have behaved much like a Wilson line and have power law divergences resulting when $\hat{\theta} = \bar{\theta}$. This divergence would result from small fluctuations of the geometry that give the geodesic infinitely small wiggles or a fractal structure, causing the distance to diverge. This does not happen here because of the restriction to two dimensions; which renders these potential divergences integrable leaving only logarithmic divergences. The remaining logarithmic divergences conspire to cancel, leaving (\ref{result1}) finite.

 An argument can be made as to why the logarithmic divergences have to cancel, leaving $L$ finite. A perturbation of the metric which changes the geometry, will result in leaving the original geodesic as a path connecting the two end points but this path will not necessarily be the shortest one. Since the new geodesic for the modified geometry will be the shortest distance between the two points, it's length must be bounded by the length of the original geodesic, which was finite. Since it must be finite, it cannot be logarithmically divergent.  It is possible, that there is some perturbative symmetry or deeper reason that causes this cancellation to happen yielding a finite result, but the author is unaware of it. It would be interesting if this cancellation of divergences would continue on in higher order terms of the quantity $L$. It is possible that a proof could be constructed for the higher order case by showing that the cancellation of higher point terms reduces to sum of repeated cancellations of the type shown here. This will have to be determined in future work.

The factor $- \frac{\b^{2}}{2}(\theta_2 - \theta_1)$  results in a $\b^{2}$ correction to the radius of the sphere. It comes from the fact that before analytic continuation to the Timelike regime, $Q = 1/b + b$\footnote{This is different in from its classical value $1/b$  because of the  requirement of the conformal weights of the primary operator $e^{2b\phi}$,  $\Delta(e^{2b\phi}) = \bar{\Delta}(e^{2b\phi}) = b(Q - b)$, and the fact that $\Delta(e^{2b\phi}) = \Bar{\Delta}(e^{2b\phi}) =1$ i.e. that $e^{2b\phi}$ transforms as $ (1,1)$ tensor so that $\int d^{2}x\sqrt{\tilde{g}}e^{2b\phi}$ is conformally invariant\cite{Ginsparg:1993is}. This conformal weight can be obtained by computing the O.P.E. of the stress tensor with the operator, $T[z]e^{2b\phi} \sim \frac{\Delta(e^{2b\phi})}{(z - w)^{2}} + \ldots$.}. This factor would be there if there was no fluctuation of the Liouville field away from the saddle point, and is independent of the gauge-fixed propagator that was derived.

\subsection{Break Down Of The Perturbation Of \texorpdfstring{$L$}{L1} For Large Separation Angle.}

\hspace{0.25in}One point of note is  that (\ref{result1}) diverges when $\theta_2 - \theta_1 \rightarrow \pi$. This is a sign that the perturbation series is breaking down, not that the distance $L$ is becoming infinite. This can be explained by noting that if the end points are taken to be the north and south poles of the sphere, the geodesic connecting them is degenerate. When the angles are not antipodal on the sphere, there is a unique unperturbed geodesic connecting them, $\varphi[\theta] = \varphi_0$, which fluctuations can be computed about. As the end points become antipodal, there are many different paths that are infinitesimally close to the true geodesic. This degeneracy means that the current expansion is not an analytic function of $\b$ as $\b\rightarrow 0$ when $\theta_2 -\theta_1 = \pi$, and hence a power series expansion around $\b\rightarrow0$ is no longer valid. This is analogous to expanding $\sqrt{x}$ around $x=0$ and noting that coefficients of the power series are infinite. The situation occurs in degenerate perturbation theory, where a perturbation breaks the degeneracy.  In the limit of the perturbation vanishing, the perturbation series begins to break down as the second order and higher terms become the same magnitude as the unperturbed states. For the present situation the result (\ref{result1}) breaks down as
\begin{equation}
(\pi - (\theta_2 - \theta_1)) \sim \frac{\hat{b}}{2\sqrt{2}\pi}\nonumber.
\end{equation}

To compute the corrections of geodesics ending on antipodal points, a resummation of series is necessary.

\subsection{The Ratio Of The Correction To The Unperturbed Distance, In The Limit Of Vanishing Angle.}

\hspace{0.25in}One other interesting fact about (\ref{result1}) is that even though the function vanishes as $\theta_2 - \theta_1 \rightarrow 0$, the ratio of the $\frac{\b^{2}}{16\pi^{2}}2\tan{\Big[\frac{\theta_2 - \theta_1}{2}\Big]}\log{|1 - \cos{[\theta_2 - \theta_1]}|}$ to the unperturbed distance $\theta_2 - \theta_1$ diverges. This is because (\ref{result1}) is not analytic at $\theta_2 - \theta_1 = 0$. This implies that (\ref{result1}) should not be trusted for very small separations of the angle. At small angles, the series breaks down as,
\be\label{planckscale}
\theta_2 - \theta_1 < \sqrt{2}e^{-\frac{8\pi^{2}}{\b^{2}}}.
\ee

Here $\sqrt{2}e^{-\frac{8\pi^{2}}{\b^{2}}}$ acts as the Planck length of the system.

 \subsection{Future Work} 

\begin{itemize}
\item \hspace{0.25in}Now that a gauge invariant propagator has been computed, many other quantities can be computed using standard techniques. Computations of the expectation value of curvature invariants or other diffeomorphism invariant quantities involving the metric, can be computed in this formalism. This can be done by expanding the Liouville factor of the metric into the saddle point contribution and the fluctuation, expanding in powers of $\b$ and using standard perturbative techniques to compute the quantity with the propagator $\big<f[\hat{\theta}]f[\bar{\theta}]\big>$ and required derivatives.

\item \hspace{0.25in}When coupling a matter CFT to Timelike Liouville theory, quantities invariant under $SL_{2}(\mathbb{C})$ transformations can be constructed. For example, The two point correlator of two fields of known scaling dimension at fixed geodesic distance.

   \hspace{0.25in} Using the results of Section \ref{dist} the correlator of two conformal matter fields of known scaling dimension at fixed geodesic distance $L$ can be computed.
\be\label{fixeddist}
<\mathcal{O}\,\mathcal{O}>  \,= \int\,d^{2}x\,d^{2}y\,\delta^{(2)}\Big[L -  \int^{y}_{x}\sqrt{g_{\mu\nu}\frac{\partial x^{\mu}}{\partial\sigma}\frac{\partial x^{\nu}}{\partial\sigma}}\,d\sigma\Big]<\mathcal{O}(x)\mathcal{O}(y)>
\ee

Here $<X(x)X(y)>$ is the correlator on the fixed reference sphere.

Na\"{i}vely there is no obstruction to extending this to $n$-point correlation functions including a delta function constraint for each pair of points, fixing there separation to a fixed physical distance.

\item \hspace{0.25in}Another quantity of note is the two point correlator of two fields of fixed scaling dimension under the influence of a probe propagator.  A scalar field under the influence of Liouville has an action of the form
\begin{align}\label{probeaction}
-\mathcal{S}_{\lambda} &= -\int d^{2}x\,\sqrt{g}\,[g^{ab}\nabla_{a}\lambda\nabla_{b}\lambda - m^{2}\lambda^{2}]\nonumber\\ &= -\int
d^{2}x\,\sqrt{\hat{g}}e^{2\b\hat{\phi}}\,[e^{-2b\hat{\phi}}\hat{g}^{ab}\nabla_{a}\lambda\nabla_{b}\lambda - m^{2}\lambda^{2}].
\end{align}

Extracting the Liouville dependence and integrating by parts, this action can be rewritten as
\begin{align}\label{probeactionuse}
-\mathcal{S}_{\lambda} &= -\int d^{2}x\,\sqrt{\hat{g}}\,[\hat{g}^{ab}\partial_{a}\lambda\partial_{b}\lambda - m^{2}e^{2\b(\hat{\phi}_{0} + f)}\lambda^{2}]\nonumber\\
&=-\int d^{2}x\,\sqrt{\hat{g}}\,[\hat{g}^{ab}\partial_{a}\lambda\partial_{b}\lambda - m^{2}e^{2\b\hat{\phi}_{0}}(1 + f + \frac{1}{2}f^{2})\lambda^{2}].
\end{align}

From (\ref{probeactionuse}) the  probe propagator and the probe Feynman rules can be obtained,

\begin{align}
<\lambda\lambda>&=\text{
\begin{fmffile}{probline}
\begin{fmfgraph*}(30,3)
\fmfleft{in}
\fmfright{out}
\fmflabel{$x$}{in}
\fmflabel{$y$}{out}
\fmf{fermion}{in,out}
\fmfdot{in,out}
\end{fmfgraph*}
\end{fmffile}}\hspace{6pt} = (\nabla^{2} + m^{2}e^{2\b\hat{\phi}_{0}})^{-1}\nonumber\\
& =\alpha_2\,\P[{\frac{1}{2}\sqrt{1+4m^{2}e^{2\b\hat{\phi}_{0}}} -1},\chi_{xy}]\nonumber\\&\hspace{0.5in}+ \alpha_3\,\Q[{\frac{1}{2}\sqrt{1+4m^{2}e^{2\b\hat{\phi}_{0}}} -1},\chi_{xy}].
\footnote{Here $\alpha_2$ and $\alpha_3$ are chosen to satisfy the boundary conditions of the field.}
\end{align}
\begin{align}&
\text{\raisebox{-0.25in}{\begin{fmffile}{propliouvert}
\begin{fmfgraph*}(20,15)
\fmfleft{i1,i2}
\fmfright{o1}
\fmf{fermion}{i1,v1}
\fmf{fermion}{i2,v1}
\fmf{dashes}{v1,o1}
\fmfdot{v1}
\fmflabel{$z$}{v1}
\end{fmfgraph*}
\end{fmffile}}} = \int\,d^{2}z\sqrt{\hat{g_z}}m^{2}e^{\hat{\phi}_0}\\
&\text{\raisebox{-0.25in}{\begin{fmffile}{propliouvert2}
\begin{fmfgraph*}(20,15)
\fmfleft{i1,i2}
\fmfright{o1,o2}
\fmf{fermion}{i1,v1}
\fmf{fermion}{i2,v1}
\fmf{dashes}{v1,o1}
\fmf{dashes}{v1,o2}
\fmfdot{v1}
\fmflabel{$z$}{v1}
\end{fmfgraph*}
\end{fmffile}}} = \frac{1}{2}\int\,d^{2}z\sqrt{\hat{g_z}}m^{2}e^{\hat{\phi}_0}.
\end{align}

One quantity that can be computed is, 
  \begin{align}\label{twopointwithprop}
  &\int d^{2}x\, d^2{}y\,\sqrt{g_x}\,\sqrt{g_y}\,<X(x)(\nabla^2 + m^2)^{-1}X(y)> \\
  &=\,\int d^{2}x\, d^{2}y\,\sqrt{\hat{g}_x}\,\sqrt{\hat{g}_y}\mathbb{Z}^{-1}\int\D\,f\, e^{f(x)}e^{f(y)}X(x)(\hat{\nabla}^2 + m^{2}e^{\phi_0}e^{f})^{-1}X(y)e^{-S_l}.
\end{align}

This is the analog of the first correction in the expansion of a Wilson line coming from a scalar mediating boson, on a fluctuating sphere. Here the coordinates, $x$, $y$, are on the reference sphere. In (\ref{twopointwithprop}) the Liouville field enters from two regimes. First from the integration measures of the coordinates
on the sphere, and second from the covariant derivative in the propagator. Taking the perturbative expansion of the Liouville field to quadratic order, (\ref{twopointwithprop}) can be
written more explicitly as
  \begin{align}\label{twopointcorr}
\int d^{2}x\, d^{2}y\,\sqrt{\hat{g}_x}\,\sqrt{\hat{g}_y}\mathbb{Z}^{-1}\int\D\,f\, [1 + &f(x) + f(y) + \frac{1}{2}f(x)^{2} + \frac{1}{2}f(y)^{2}
+f(x)f(y)]\times\nonumber\\&\times\chi(x)(\hat{\nabla}^2 + m^{2}(1 + f + \frac{1}{2}f^2))^{-1}\chi(y)e^{-S_l}.
  \end{align}
  
  where $S_l$ is the the linearized gauge fixed action Liouville action. The remaining Feynman rules obtained from (\ref{twopointcorr}) are,
\begin{align}
&\text{\raisebox{-0.25in}{\begin{fmffile}{3matterprobliouvert}
\begin{fmfgraph*}(20,15)
\fmfleft{i1,i2}
\fmfright{o1}
\fmf{fermion}{i1,v1}
\fmf{wiggly}{i2,v1}
\fmf{dashes}{v1,o1}
\fmfdot{v1}
\fmflabel{$z$}{v1}
\end{fmfgraph*}
\end{fmffile}}} = \int\,d^{z}\sqrt{\hat{g}_z}\hspace{20pt}
\text{\raisebox{-0.25in}{\begin{fmffile}{4matterprobliouvert}
\begin{fmfgraph*}(20,15)
\fmfleft{i1,i2}
\fmfright{o1,o2}
\fmf{fermion}{i1,v1}
\fmf{wiggly}{i2,v1}
\fmf{dashes}{v1,o1}
\fmf{dashes}{v1,o2}
\fmfdot{v1}
\fmflabel{$z$}{v1}
\end{fmfgraph*}
\end{fmffile}}} = \frac{1}{2}\int\,d^{z}\sqrt{\hat{g}_z}\\&
\hspace{1in}\text{\begin{fmffile}{matterprop}
\begin{fmfgraph*}(25,3)
\fmfleft{in}
\fmfright{out}
\fmflabel{$x$}{in}
\fmflabel{$y$}{out}
\fmf{wiggly}{in,out}
\fmfdot{in,out}
\end{fmfgraph*}
\end{fmffile}} \hspace{10pt}= \tilde{C}_{xy}(1 - \chi_{xy})^{-2\Delta}.
\end{align}

It follows that (\ref{twopointcorr}) corresponds to the Feynman diagrams in Figure \{\ref{one}\}.
\begin{figure}[ht]\label{one}
\raisebox{-0.55in}{\begin{fmffile}{bubble1}
\begin{fmfgraph*}(30,30)
\fmfleft{x}
\fmfright{y}
\fmflabel{$x$}{v1}
\fmflabel{$y$}{v2}
\fmf{phantom,tension=4}{x,v1}
\fmf{phantom,tension=4}{v2,y}
\fmf{wiggly,left,tension=0.4}{v1,v2}
\fmf{fermion,left,tension=-0.4}{v2,v1}
\fmf{phantom}{v1,v2}
\fmfdot{v1,v2}
\end{fmfgraph*}
\end{fmffile}}
$+$ \hspace{-0.2in}
\raisebox{-0.63in}{
\begin{fmffile}{bubble2}
\begin{fmfgraph*}(30,30)
\fmfleft{x}
\fmfright{y}
\fmftop{w,z}
\fmflabel{$x$}{v1}
\fmflabel{$y$}{v2}
\fmflabel{$z$}{w1}
\fmflabel{$w$}{w2}
\fmf{phantom,tension=2}{z,w1}
\fmf{phantom,tension=2}{w2,w}
\fmf{phantom,tension=4.5}{x,v1}
\fmf{phantom,tension=4.5}{v2,y}
\fmf{wiggly,left,tension=-5.9}{v1,v2}
\fmf{fermion,right=0.2}{v1,w1}
\fmf{fermion,right=0.2}{w1,w2}
\fmf{fermion,right=0.2}{w2,v2}
\fmf{dashes,right,tension=2}{w1,w2}
\fmfdot{v1,v2,w1,w2}
\end{fmfgraph*}
\end{fmffile}}
$+$\raisebox{-0.63in}{
\begin{fmffile}{bubble3}
\begin{fmfgraph*}(30,30)
\fmfleft{x}
\fmfright{y}
\fmftop{z}
\fmftop{w}
\fmflabel{$x$}{v1}
\fmflabel{$y$}{v2}
\fmflabel{$z$}{w1}
\fmf{phantom,tension=2}{z,w1}
\fmf{phantom,tension=5}{x,v1}
\fmf{phantom,tension=5}{v2,y}
\fmf{wiggly,left,tension=-6.9}{v1,v2}
\fmf{fermion,right=0.2}{v1,w1}
\fmf{fermion,right=0.2}{w1,v2}
\fmf{dashes,right,tension=0.5}{w1,w}
\fmf{dashes,right,tension=0.5}{w,w1}
\fmfdot{v1,v2,w1}
\end{fmfgraph*}
\end{fmffile}}\hspace{0.1in}$+$
\raisebox{-0.69in}{
\begin{fmffile}{bubble4}
\begin{fmfgraph*}(30,30)
\fmfleft{x}
\fmfright{y}
\fmftop{z}
\fmfbottom{w}
\fmflabel{$x$}{v1}
\fmflabel{$y$}{v2}
\fmflabel{$z$}{w1}
\fmf{phantom,tension=2}{z,w1}
\fmf{phantom,tension=4.5}{x,v1}
\fmf{phantom,tension=5}{v2,y}
\fmf{phantom,tension=-0.6}{v2,w}
\fmf{wiggly,left,tension=-6.9}{v1,v2}
\fmf{fermion,right=0.2}{v1,w1}
\fmf{fermion,right=0.2}{w1,v2}
\fmf{dashes,left=1}{w1,v1}
\fmfdot{v1,v2,w1}
\end{fmfgraph*}
\end{fmffile}}

$+$\hspace{-0.2in}
\raisebox{-0.66in}{
\begin{fmffile}{bubble5}
\begin{fmfgraph*}(30,30)
\fmfleft{x}
\fmfright{y}
\fmftop{z}
\fmfbottom{w}
\fmflabel{$x$}{v1}
\fmflabel{$y$}{v2}
\fmflabel{$z$}{w1}
\fmf{phantom,tension=2}{z,w1}
\fmf{phantom,tension=5.5}{x,v1}
\fmf{phantom,tension=5.5}{v2,y}
\fmf{phantom,tension=-0.3}{v1,w}
\fmf{wiggly,left,tension=-6.9}{v1,v2}
\fmf{fermion,right=0.2}{v1,w1}
\fmf{fermion,right=0.2}{w1,v2}
\fmf{dashes,right=1}{w1,v2}
\fmfdot{v1,v2,w1}
\end{fmfgraph*}
\end{fmffile}}
$+$
\raisebox{-0.63in}{
\begin{fmffile}{bubble6}
\begin{fmfgraph*}(30,30)
\fmfleft{x}
\fmfright{y}
\fmflabel{$x$}{v1}
\fmflabel{$y$}{v2}
\fmf{phantom,tension=5}{x,v1}
\fmf{phantom,tension=5}{v2,y}
\fmf{wiggly,left,tension=-6.9}{v1,v2}
\fmf{fermion,right=0.2}{v1,v2}
\fmf{dashes,tension=2.5}{v1,v1}
\fmfdot{v1,v2}
\end{fmfgraph*}
\end{fmffile}}$+$
\raisebox{-0.63in}{
\begin{fmffile}{bubble7}
\begin{fmfgraph*}(30,30)
\fmfleft{x}
\fmfright{y}
\fmflabel{$x$}{v1}
\fmflabel{$y$}{v2}
\fmf{phantom,tension=5}{x,v1}
\fmf{phantom,tension=5}{v2,y}
\fmf{wiggly,left,tension=-6.9}{v1,v2}
\fmf{fermion,right=0.2}{v1,v2}
\fmf{dashes,top,tension=2.5}{v2,v2}
\fmfdot{v1,v2}
\end{fmfgraph*}
\end{fmffile}}$+$
\raisebox{-0.535in}{
\begin{fmffile}{bubble8}
\begin{fmfgraph*}(30,30)
\fmfleft{x}
\fmfright{y}
\fmflabel{$x$}{v1}
\fmflabel{$y$}{v2}
\fmf{phantom,tension=8}{x,v1}
\fmf{phantom,tension=8}{v2,y}
\fmf{dashes}{v1,v2}
\fmf{wiggly,left,tension=2}{v1,v2}
\fmf{fermion,left,tension=-2}{v2,v1}
\fmf{phantom}{v1,v2}
\fmfdot{v1,v2}
\end{fmfgraph*}
\end{fmffile}}
\caption{Feynman diagrams corresponding to (\ref{twopointcorr}). The third, sixth, and seventh diagrams correspond to renormalization of the Liouville coupling to the conformal field and
the probe mass respectively.}
\end{figure}
\item \hspace{0.25in}Lastly, there is the question of if there is a Gribov Ambiguity with constraint used to address the gauge redundancy. The current work has fixed the gauge of the $SL_{2}(\mathbb{C})$  transformations locally, but there is no guarantee that it has been fixed globally, so it is still possible that there is a non-perturbative failure of B.R.S.T. symmetry after following this procedure, i.e. the gauge fixing of the dipole may not be the unique way to fix the gauge\cite{1978NuPhB.139....1G}. A possible avenue forward could be to construct a proof showing that the fixing of the dipole is a unique gauge condition or determine whether a Gribov Ambiguity occurs.
%
\end{itemize}
\section{Acknowledgements}
\hspace{0.25in}The author would like to thank Leonard Susskind for suggesting the problem. Discussions with Dionysis Anninos, Shamik Banerjee,  Daniel Harlow, Shamit Kachru, Victoria Martin, Edgar Shaghoulian, Stephen Shenker, Douglas Stanford, Leonard Susskind, and Edward Witten have been very useful in the course of this project. The Author would also like to thank The Cosmology and Complexity conference at Hydra for their hospitality while this work was in progress. The research of J.M. is supported by NSF grant 0756174 and the Stanford Institute for Theoretical Physics. The research of the author is also partially supported by the Mellam Family Fellowship and the Stanford School of Humanities and Sciences Fellowship.
\newpage
\bibliographystyle{JHEP}
\bibliography{bibliography}
\end{document}